\documentclass[aps,preprint,showpacs,nofootinbib,12pt]{revtex4}
\usepackage{graphicx}% Include figure files
\usepackage{epsfig}
\usepackage[dvips]{color}

\begin{document}

\begin{center}
{\Large {\bf Search for $B_c(ns)$ via the $B_c(ns)\to
B_c(ms)\pi^+\pi^-$ transition at LHCb and $Z_0$ factory}}
\end{center}

\vspace*{0.3cm}
\begin{center}
{Hong-Wei Ke$^1$ and Xue-Qian Li$^2$  }\\
\vspace*{0.3cm} {\it 1. School of Science, Tianjin University,
Tianjin, 300072, China\\

2. School of Physics, Nankai University, Tianjin 300071, China}\\

\end{center}

\vspace*{0.5cm}

\begin{center}
\begin{minipage}{12cm}

\noindent Abstract:\\
 \noindent \vspace*{0.5cm}
It is interesting to study the characteristics of the whole family
of $B_c$ which contains two different heavy flavors. LHC and the
proposed $Z^0$ factory provide an opportunity because a large
database on the $B_c$ family will be achieved. $B_c$ and its excited
states can be identified via their decay modes. As suggested by
experimentalists, $B_c^*(ns)\to B_c+\gamma$ is not easy to be
clearly measured, instead, the trajectories of $\pi^+$ and $\pi^-$
occurring in the decay of $B_c(ns)\to B_c(ms)+\pi^+\pi^-$ ( $n>m$ )
can be unambiguously identified, thus the measurement seems easier
and more reliable, therefore this mode is more favorable at early
running stage of LHCb and the proposed $Z^0$ factory. In this work,
we calculate the rate of $B_c(ns)\to B_c(ms)+\pi^+\pi^-$ in terms of
the QCD multipole-expansion and the numerical results indicate that
the experimental measurements with the luminosity of LHC and $Z^0$
factory  are feasible.

 PACS number(s): 13.25.-k
\end{minipage}

\end{center}

\baselineskip 22pt
\newpage

\section{Introduction}

The heavy quarkonia such as charmonia and bottomia have been
experimentally and theoretically explored for several decades
already because large database on them is available. In comparison,
the physics on $B_c$ has not been thoroughly studied yet. The reason
is obvious that $B_c$ contains two different heavy flavors, so
unlike quarkona, it is produced via more suppressed processes at
$e^+e^-$ colliders. The earlier work \cite{Chang} indicates that at
the luminosity of regular $e^+e^-$ colliders (the LEP I and II), one
cannot expect to observe $B_c$ production, i.e. its production rate
is too small to be measured. Therefore, people should turn to hadron
colliders. As predicted \cite{Chang}, $B_c$ was observed at
TEVATRON\cite{Cheung:1999ir} a while ago. At the energy and
luminosity of LHC, one may anticipate  $B_c$ events thousand times
more than at TEVATRON. Now a project of constructing a $Z^0$ factory
is proposed which will provide sufficiently high luminosity at the
$Z^0$ pole. Even though the production process $e^+e^-\to
B_c+\overline {B_c}$ where a pair of heavy quarks ($c\bar c$ or
$b\bar b$) emerges from a hard gluon emission, is suppressed, the
high luminosity and the pole effect would greatly enhance the
production rate, i.e. the enhancement compensates the suppression
and enables the production of $B_c$ measurable.

Since $B_c$ is made of two different heavy flavors, its decay
characteristics would somehow distinct from the heavy quarkonia
which contains a pair of heavy quarks of the same flavor. Namely,
the two constituents in quarkonia annihilate into gluons which
afterwards hadronize.  Although this mode is OZI suppressed, it is a
strong interaction-induced process and has a larger decay width.
Nevertheless the ground state of $B_c$ family can only decay via
weak interaction and its lifetime has been carefully
studied\cite{Chang1}. It is interesting to investigate such mesons
and we are not only interested in the ground state, but also its
excited states. The lowest excited states would be the vector
$B_c^*\; (1s)$ and pseudoscalar $B_c(2s)\; (0^-)$ and the latter one
is the first radially excited state of the family. A simple analysis
estimates the splitting between $B_c(1s)$ and $B^*_c(1s)$ is about
50 to 80 MeV, so that $B_C^*\to B_c+\pi^0$ is forbidden by the
energy-conservation and the hadronic decays of $B_c^*$ can only
occur via weak processes which are obviously suppressed. Only
possible transition is the radiative decay $B_c^*\to B_c+\gamma$,
but at LHC, detection of a single photon from a messy background is
very difficult. Instead, in the decay $B_c(ns)\to
B_c(ms)+\pi^+\pi^-$ the daughter charged pions are very easy to be
identified. Therefore, our experimental colleagues strongly suggest
us to investigate the channel $B_c(ns)\to B_c(ms)+\pi^+\pi^-$.
Definitely, we can gain more information about the $B_c$ family.

Since $B_c$ is composed of two heavy quarks, the relativistic
effects may not be too serious, thus the potential model can be a
good choice for determining the spectra of $B_c$ and its excited
states. In analog to dealing with heavy charmonia and bottomia, we
employ the Cornell potential to calculate the masses of $B_c(ns)$
and $B_c^*(ns)$  where the mass of the ground state $B_c(1s)$ is
taken as input for fixing concerned model parameters.

Following literature\cite{Gottfried,YK1,K2,Y1,K1} we evaluate the
decay rate of $B_c(ns)\to B_c(ms)+\pi\pi$ in terms of the QCD
multipole-expansion. The picture is depicted as that the initial
$B_c$ transits into a hybrid state $b\bar cg$ where $b\bar c$
stays in a color-octet, by emitting a gluon, and then the hybrid
turns into $B_c(1s)$ by emitting the second gluon. The two gluons
eventually hadronize into two pions. In the transitions
$B_c(ns)\rightarrow B_c(ms)+\pi+\pi$, the momentum transfer is not
large and the perturbative method does not apply. The  QCD
multipole expansion (QCDME) method suggested by Gottfried, Yan and
Kuang\cite{Gottfried,YK1,K2,Y1,K1} properly treats the light-meson
emission process. In the picture of the QCD multipole expansion,
the two emitted gluons are not energetic particles, but described
by a chromo filed of TM or TE modes\cite{soft}. It is worth
emphasizing again that the two gluons are not free gluons in the
sense of the perturbative quantum field theory, but a field in the
QCD multipole expansion. It is easy to understand that such
transition is dominated by the E1-E1 mode, while the M1-M1 mode is
suppressed for the heavy quarkonia case.

The two chromo-E1 transitions are dealt with in the regular
framework of multipole-expansion. The key point is how to properly
obtain the mass-spectrum and wavefunction of the intermediate hybrid
state. Isgur and Paton\cite{Isgur} suggested to use a modified
potential to describe the hybrid states, but there are a few free
parameters to be determined. Before, the parameter in the
potential\cite{BT} which Kuang and Yan employed, was fixed by
assuming the $\psi(4040)$ to be the hybrid \cite{YK1}. Thanks to the
achievements of BELLE, CLEO and BES, an abundant database on such
two-pion emission processes has been available. In our previous work
\cite{Ke2007}, we carefully discussed the cases of
$\Upsilon(ns)\rightarrow \Upsilon (ms)+\pi+\pi$ and
$\psi(ns)\rightarrow \psi(ms)+\pi+\pi$ in terms of the QCD
multipole-expansion, then by the method of minimizing $\chi^2$ which
is widely adopted in analyzing experimental data, we eventually fix
the parameters in the potential. For the hybrid state we used three
different potential models\cite{Isgur,Swanson,Allen} and noticed
that the potential form proposed by Allen $et\, al.$\cite{Allen}
better coincides with the lattice result. According to the the
potential parameters gained by fitting the spectra of heavy
quarkonia $\Upsilon$ and $\psi$ families, one can estimate the
parameters for the members of the $B_c$ family,  by slightly varying
the values of corresponding parameters (see below for details).

In this work, we apply the QCD multipole expansion method \cite{YK1}
and the potential form suggested by Allen et al. \cite{Allen} to
calculate the transition rate of $B_c(ns)\rightarrow B_c(ms)+\pi\pi$
where the potential parameters for the hybrids $|b\bar cg>(|b\bar c
g>)$ are the same as that we obtained for $|b\bar bg>$ and $|c\bar
cg>$. Since the spectra and wavefunctions of higher exited states of
$B_c$ are even harder to be accurately derived, at this stage, we
only concern the transitions from the radially excited states
$B_c(3s)$ and $B_c(2s)$ into the lower states via emitting two
pions.

The derivation and numerical computations are straightforward and
very similar to the procedures carried out in literature,
therefore, below unless necessary for clarity, we omit some
technical details. This work is organized as follows. After the
introduction, we present the theoretical formulae for the
transition $B_c(ns)\to B_c(ms)+\pi\pi$, and then in sec.III, we
list our numerical results along with all input parameters in
tables, the last section is devoted to our discussion and
conclusion.

%Our numerical results indicate that the ground states of pure
%hybrid $|b\bar cg>$ is about $7.63\sim7.64$GeV for the
%spin-singlet $b\bar c$ in hybrid.

\begin{figure}[!h]\label{p1}
\begin{center}
%\begin{tabular}{ccc}
\scalebox{0.8}{\includegraphics{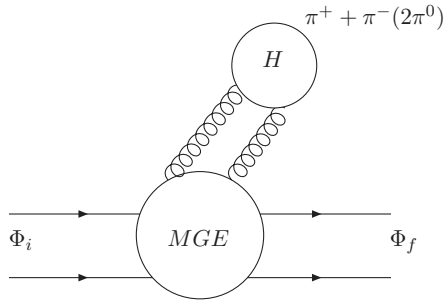}}
%\end{tabular}
\end{center}
\caption{The two-gluon emission diagram and the two  gluons
eventually hadronize into two pions. The intermediate state is a
hybrid state $|b\bar cg>$. }
\end{figure}

\section{Formulation}
\subsection{The transition width }
The theoretical framework about the QCD Multiploe Expansion method
is well framed in Refs\cite{YK1,K2,Y1,K1}, and all the concerned
formulas are presented in those papers. Here we only copy a few
formulas which are necessary for evaluating the widths of
$B_c(ns)\rightarrow B_c(ms)+\pi+\pi$ $(n>m)$ in this work. For all
details, the readers are suggested to refer to the original works
and references therein. The general formula for the rate caused by
double E1 transitions was given in Ref.\cite{YK1} as
\begin{eqnarray}\label{s0}
\Gamma
&=&\delta_{l_il_f}\delta_{J_iJ_f}(G|C_1|^2-\frac{2}{3}H|C_2|^2)\,\,|\sum_l(2l+1)
\left ( \begin{array}{ccc}
  l_i& 1 &l \\
  0 & 0 & 0 \end{array} \right )\left ( \begin{array}{ccc}
  l& 1 &l_i \\
  0 & 0 & 0 \end{array} \right
  )f^{l}_{i,f}|^2\nonumber\\&&+(2l_i+1)(2l_f+1)(2J_f+1)
  \sum_k(2k+1)[1+(-1)^k]
  \left \{ \begin{array}{ccc}
  s& l_f &J_f \\
  k & J_i & l_i \end{array} \right \}H|C_2|^2\nonumber\\&&
  |\sum_l(2l+1)
\left ( \begin{array}{ccc}
  l_f& 1 &l \\
  0 & 0 & 0 \end{array} \right )\left ( \begin{array}{ccc}
  l& 1 &l_i \\
  0 & 0 & 0 \end{array} \right
  )
\left \{ \begin{array}{ccc}
  l_i& l &1 \\
  1 & k & l_f \end{array} \right
  \}
  f^{l}_{i,f}|^2,
\end{eqnarray}
where $l_{i(f)}$ and $J_{i(f)}$ are orbital and total angular
momenta of the initial( final) heavy quarkonia  and the total spin
of the initial and final states respectively. $l$ is the angular
momentum of the color-octet intermediate state.
 $|C_1|^2$ and $|C_2|^2$ are  constants to be determined which come from the
hadronization of gluons into pions. $G$ and $H$ are the
phase-space integrals  whose concrete forms were given in
\cite{YK1,K2}. Obviously the first term corresponds to an S-wave
and the second term to a mixing of S and D-waves.
 %as
%\begin{eqnarray}\label{s2}
%f^{l,P_{I},P_{F}}_{n_{I},l_{I},n_{F},l_{F}}=\sum_K\frac{\int
%R_F(r)r^{P_F}R^*_{Kl}(r)r^2dr \int
%R^*_{Kl}(r')r'^{P_I}R_I(r')r'^2dr'}{M_I-E_{Kl}},
%\end{eqnarray}
%where $n_{I},n_{F}$ are the principal quantum numbers of initial
%and final states, $l_{I},l_{F}$ are the angular momenta of the
%initial and final states, , $P_{I},P_{F}$ are the indices related
%to the multipole radiation, for the E1 radiation $P_{I},P_{F}$=1
%and $l=1$. $R_I,R_F$ and $R_{Kl}$ are the radial wave functions of
%the initial and final states, $M_I$ is the mass of initial
%quarkonium and $E_{Kl}$ is the energy eigenvalue of the
%intermediate hybrid state.

In terms of  Eq.(\ref{s0}) the transition rate of a pseudoscalar
meson into another pseudoscalar meson with a two-pion emission can
be written as
\begin{eqnarray}\label{s1}
\Gamma(n_{i}{}^{1}S_{0}\rightarrow
n_{f}{}^{1}S_{0})=|C_1|^2G|f^{l}_{if}|^2,
\end{eqnarray}
which is similar to that between two
vector-quarkonia\cite{YK1,K2}. $n_{i},n_{f}$ are the principal
quantum numbers of initial and final states and $f^{l}_{if}$ is
the overlapping integration over the concerned hadronic wave
functions,
\begin{eqnarray}\label{s2}
f^{l}_{i,f}=\sum_K\frac{\int R_f(r)r^{P_f}R^*_{Kl}(r)r^2dr \int
R^*_{Kl}(r')r'^{P_i}R_i(r')r'^2dr'}{M_i-E_{Kl}},
\end{eqnarray}
where  $P_{i},\ P_{f}$ are the indices related to the multipole
radiation, for the E1 radiation $P_{i},P_{f}$=1 and $l=1$. $R_i,R_f$
and $R_{Kl}$ are the radial wave functions of the initial and final
states, $M_i$ is the mass of initial quarkonium and $E_{Kl}$ is the
energy eigenvalue of the intermediate hybrid state. The sum over the
principle number of the intermediate hybrid state $K$ is truncated
at $K=7$ because the contributions from higher excited states are
too small and can be safely neglected.

Because strong interaction is blind to flavor and electrical charge,
the whole scenario should be applicable for the $B_c$ cases.

\subsection{The phenomenological potential}

In this study we employ generalized Cornell
potential\cite{cornell} which includes a spin-related term
\cite{ss} for the initial and final states
 as
% The potential reads as
%\begin{eqnarray}\label{cornell}
%V(r)=-\frac{\kappa}{r}+br,
%\end{eqnarray}
\begin{eqnarray}\label{s4}
V(r)=-\frac{\kappa}{r}+br+V_{s}(r)+V_0,
\end{eqnarray}
where $\kappa=\frac{4\alpha_s(r)}{3}$ and the coupling $\alpha_s(r)$
can be treated as a phenomenological constant while calculating the
spectra of quarkonia. Thus, for the phenomenological application,
one does not need to consider their QCD running. The spin-related
term $V_{s}$ is,
\begin{eqnarray}\label{spin}
&&V_{s}=\frac{8\pi\kappa}{3m_Q^2}\delta_\sigma(r)\overrightarrow{S}_Q
\cdot\overrightarrow{S}_{\bar{Q'}},
\end{eqnarray}
with
\begin{eqnarray}\label{spin1}
\delta_\sigma(r)=(\frac{\sigma}{\sqrt{\pi}})^3e^{-\sigma^2r^2},
\end{eqnarray}
and $V_{0}$ is the zero-point energy\cite{Ke2007}.

%\subsection{The potential for hybrids}
The intermediate state between the two gluon-emissions is a hybrid
state, namely the $Q\bar Q'$ resides in a color-octet. It was
indicated that one still can use an effective potential to describe
the color-octet $Q\bar Q'$ state \cite{BT,Isgur,Swanson,Allen}. In
literature, there are four different effective potential forms which
are respectively suggested by the authors of
Refs.\cite{BT,Isgur,Swanson,Allen}. In our earlier work
\cite{Ke2007} we employed  the three models of them to study the
hybrid state $|Q\bar Q'g>$ ($Q$ stands as $b$ or $c$ in
\cite{Ke2007}) whereas, Yan and Kuang \cite{YK1} used the potential
given by Buchm\"uller and Tye \cite{BT}. We find that the potential
form suggested by Allen $et al.$\cite{Allen} coincides with the
lattice result better than the others, thus we will use that
potential in this work.

The corresponding potential form is
\begin{eqnarray}\label{s5}
V_i(r)=\frac{\kappa'}{8}+\sqrt{(b'r)^2+2\pi b'}+V_0'.
\end{eqnarray}

Because  the authors of Refs.\cite{Isgur,Swanson,Allen} did not
consider the spin-related term, we have modified the potential by
adding the spin-related term $V_s$ which has the same form given
above in Eq.(\ref{spin}), then the potential becomes:
\begin{eqnarray}\label{s6}
V(r)=V_i+V_s.
\end{eqnarray}
With this modification, we can investigate the spin-splitting
effects.

\section{Numerical results}

At first we need to fix the  parameters in the potentials. In our
earlier work\cite{Ke2007},  we re-fitted the spectra of the
quarkonia to obtain the corresponding potential parameters in
Eq.(\ref{s4}). The values of the parameters  are listed in Table
\ref{t1}.

\begin{table}[!h]
\caption{The potential parameters for $c\bar{c}$ and $b\bar b$}
\begin{center}
\begin{tabular}{|c|c|c|c|c|c|c|}
  % after \\: \hline or \cline{col1-col2} \cline{col3-col4} ...
  \hline
 &$\kappa$ &$b$(GeV$^2$) &$m$(GeV)& $\sigma$(GeV$^2)$&$V_{0}$(GeV)\\\hline
   $c\bar c$ & 0.67& 0.16 & 1.78 &1.6  &-0.6\\\hline
     $b\bar b$    & 0.53& 0.16 & 5.13 &1.7  &-0.6\\\hline
\end{tabular}\label{t1}
\end{center}
\end{table}

One can see that the parameters $b$ and $V_0$ in Eq.(\ref{s4}) are
the same for $b\bar b$ and $c\bar c$, so that we suppose they are
unchanged for $b\bar c$. Since  the difference between the values
of the parameter $\sigma$ for $b\bar b$ and $c\bar c$ is small, it
is plausible to choose $\sigma=1.65$ for $b\bar c \,(\bar b c)$.
In the calculation for quarkonia $m_Q$ is the mass of the
constituent quark (b or c), instead, for $b\bar c$ one should use
the reduced mass. Here we set $m_c=1.78$ GeV and $m_b=5.13$ GeV
for concrete numerical computations. Fitting the mass of
$B_c(0^-)$  $6.276$ GeV \cite{PDG}, we  obtain $\kappa=0.58$. Then
with the given potential we predict the masses of a few other
$b\bar c$ states listed in the following table.

\begin{table}[!h]
\caption{ our prediction on mass of some  $b\bar c$ mesons}
\begin{center}
\begin{tabular}{|c|c|c|c|c|c|c|}
  % after \\: \hline or \cline{col1-col2} \cline{col3-col4} ...
  \hline
 &$1s$ &$2s$ &$3s$\\\hline
   $0^-$(GeV) & 6.276&6.880  &7.254 \\\hline
     $1^-$ (GeV)   &6.356 & 6.908 & 7.274 \\\hline
\end{tabular}\label{t2}
\end{center}
\end{table}

For the hybrid potential (\ref{s5}), the strategy in our earlier
work \cite{Ke2007} is that we use the minimal $\chi^2$ method which
is widely adopted in analysis of experimental data, to determine the
potential parameters which are listed in Table \ref{t3}. One can see
that $b'$ and $V_0'$ in Eq.(\ref{s5}) are the same but $\kappa'$ is
different for $b\bar b g$ and $c\bar c g$. Because no direct
measurements on transitions $B_c^{(*)}(ns)\rightarrow
B_c^{(*)}(ms)+\pi+\pi$ have ever been conducted so far, we cannot
determine the parameter $\kappa'$ for $b\bar c g$ in terms of
available data as we did for the heavy quarkonia. However one can
expect that $\kappa'$ should fall in the region between the values
for the $b\bar bg$ and $c\bar cg$ systems, thus we will vary this
value slightly within the range to study a possible dependence of
the numerical results on $\kappa'$. The dependence is shown in Table
\ref{t4} (see below).

\begin{table}[!h]
\caption{ potential parameters for $c\bar{c}g$ and $b\bar bg$}
\begin{center}
\begin{tabular}{|c|c|c|c|c|c|c|}
  % after \\: \hline or \cline{col1-col2} \cline{col3-col4} ...
  \hline
 &$\kappa'$ &$b'$(GeV$^2$) &$m$(GeV)& $V'_{0}$(GeV)\\\hline
   $c\bar cg$ & 0.54& 0.24 & 1.78   &-0.8\\\hline
     $b\bar bg$    & 0.40& 0.24 & 5.13   &-0.8\\\hline
\end{tabular}\label{t3}
\end{center}
\end{table}

%According to the measured value for $\Gamma(\psi(2S)\rightarrow
%J/\psi\,\pi\,\pi)$:
%\begin{eqnarray*}\label{s9}
%&&\Gamma_{tot}(\psi(2S))=337\pm13 \rm{keV} \\&&
%B(\psi(2S)\rightarrow J/\psi\pi^+\pi^-)=(31.8\pm0.6)\% \\&&
%B(\psi(2S)\rightarrow J/\psi\pi^0\pi^0)=(16.46\pm0.35)\%
%\end{eqnarray*}
% We obtain
%\begin{eqnarray*}\label{s10}
%|C_1|^2= 182.12\times10^{-6}.\qquad
%\end{eqnarray*}

Then there is still a free parameter $|C_1|^2$ in the decay rate
Eq.(\ref{s1}). It is noted that $C_1^2$ is a factor related to the
hadronization of gluons into two pions, so should be universal for
$\psi$, $\Upsilon$ and $B_c$ meson decays. Because $C_1$ is fully
determined by the non-perturbative QCD effects, it cannot be
derived from an underlying principle so far. Thus in
Ref.\cite{Ke2007} we fixed $|C_1|^2= 182.12\times10^{-6}$ in terms
of the well measured decay width $\Gamma(\psi(2S)\rightarrow
J/\psi+\pi+\pi)$.

With these potential parameters, we solve the Schr\"odinger
equations to obtain wave functions  and masses of the initial,
final and intermediate states which appear in the overlapping
integration $f^{l}_{if}$. Then we can compute the corresponding
widths of the concerned modes. As indicated above, in our
calculation we vary $\kappa'$ in a small region from 0.40
($\kappa'(c\bar c g)$) to 0.54 ($\kappa'(b\bar b g)$). The
corresponding results can be seen in Table\ref{t4}. For these
numerical results, we observe that  $\Gamma(B_c(3S)\rightarrow
B_c(2S))\pi\pi$ and $\Gamma(B_c(2S)\rightarrow B_c(1S))\pi\pi$ are
not very sensitive to the value of $\kappa'$, but
$\Gamma(B_c(3S)\rightarrow B_c(1S))\pi\pi$ is.

\begin{table}[!h]
\caption{prediction at $\sigma=1.65$(widths in units of KeV)}
\begin{center}
\begin{tabular}{|c|c|c|c|c|c|c|c|c|c|c|}
  % after \\: \hline or \cline{col1-col2} \cline{col3-col4} ...
  \hline
 $\kappa'$ & 0.40&0.42&0.44&0.46&0.48&0.50&0.52&0.54\\\hline
 $M_{b\bar c g}$(GeV)($l=0,K=1$)&7.629& 7.630&7.631&7.632&7.633&  7.634&7.635&7.636\\
$M_{b\bar c g}$(GeV)($l=1,K=1$)&7.800& 7.800&7.801&7.802&7.803&  7.804&7.805&7.806\\
$M_{b\bar c g}$(GeV)($l=1,K=2$)&8.118& 8.119&8.120&8.121&8.121&  8.122&8.123&8.123\\
$M_{b\bar c g}$(GeV)($l=1,K=3$)&8.417& 8.418&8.418&8.419&8.420&  8.420&8.421&8.422\\
$M_{b\bar c g}$(GeV)($l=1,K=4$)&8.700& 8.701&8.701&8.702&8.703&  8.703&8.704&8.704\\
$M_{b\bar c g}$(GeV)($l=1,K=5$)&8.970& 8.971&8.971&8.972&8.972& 8.9735&8.973&8.974\\
$M_{b\bar c g}$(GeV)($l=1,K=6$)&9.230&9.230&9.230&9.231&9.231&9.232&9.232&9.233\\
$M_{b\bar c g}$(GeV)($l=1,K=7$)&9.479& 9.480&9.480&9.481&9.481&9.481&9.482&9.483\\
   $\Gamma(B_c(3S)\rightarrow B_c(2S))\pi\pi$& 11.01&11.07&11.14&10.90&10.95&10.99&11.06&11.11  \\
  $\Gamma(B_c(3S)\rightarrow B_c(1S))\pi\pi$& 4.91&3.95&2.98&5.64&4.85&4.01&3.05&2.38  \\
   $\Gamma(B_c(2S)\rightarrow B_c(1S))\pi\pi$& 64.13&64.01&63.90&63.71&63.60&63.48&63.37&63.25  \\  \hline
\end{tabular}\label{t4}
\end{center}
\end{table}

%\begin{table}[!h]
%$\caption{prediction at $\sigma=1.7$(widths in units of KeV)}
%\begin{center}
%\begin{tabular}{|c|c|c|c|c|c|c|c|c|c|c|}
  % after \\: \hline or \cline{col1-col2} \cline{col3-col4} ...
%  \hline
% $\kappa'$ & 0.40&0.42&0.44&0.46&0.48&0.50&0.52&0.54\\\hline
%$M_{b\bar c g}$(GeV)&7.629& 7.630&7.631&7.632&7.633&  7.634&7.635&7.636\\
%   $\Gamma(B_c(3S)\rightarrow B_c(2S))\pi\pi$& 11.30&11.38&11.43&11.20&11.25&11.32&11.36&11.13  \\
%  $\Gamma(B_c(3S)\rightarrow B_c(1S))\pi\pi$& 2.78&1.91&1.29&3.23&2.68&1.86&1.28&3.33  \\
%   $\Gamma(B_c(2S)\rightarrow B_c(1S))\pi\pi$& 65.41&65.31&65.15&64.99&64.87&64.76&64.60&64.45  \\  \hline
%\end{tabular}\label{t5}
%\end{center}
%\end{table}

\section{Our conclusion and discussion}

Study on $B_c$, $B_c^*$ mesons and their radial and angular excited
states is important because they are the last heavy mesons and are
composed of two different heavy flavors. Because of their special
structures, a thorough study on the production and decay processes
where $B_c$ and its excited states are involved may shed more light
on the fundamental interactions, especially the non-perturbative
QCD, moreover, may provide some hints to new physics beyond the
standard model. Therefore, this field attracts attentions of
theorists and experimentalists of high energy physics. The main
obstacle for the study is that the production rate of $B_c$ is small
as Chang and his collaborators indicated \cite{Chang}. However, as
LHC begins running, the high luminosity would provide sufficiently
large data sample, moreover, a proposed Z-factory with a luminosity
much higher than the LEP-I, would offer a clean environment for the
$B_c$ research.

Among all the decay modes, $B_c(ns)\rightarrow B_c(ms)+\pi+\pi\;
(n>m)$ is a favorable one for investigating the $B_c$ family because
it is a strong-interaction process, moreover, it is also an ideal
place to study the heavy hybrids $|b\bar cg>(|\bar bcg>)$. Of
course, the radiative decay $B_c^*\rightarrow B_c+\gamma$ is also a
place to study the family \cite{Choi2009}, but as our experimental
colleagues suggest, at LHCb, the $\gamma$ detection would be
difficult. Instead, the two charged pions are easy to be identified
at LHCb detector. A rough estimate of the mass of $B_c^*$ in terms
of the potential model indicates that the mass of $B_c^*$ is 6.36
GeV which is close to that given in Ref. \cite{Colangelo2000}. It
only 80 MeV heavier than the ground state $B_c$, thus the mode
$B_c^*\rightarrow B_c+\pi$ is forbidden by the final phase space.

For the decay $B_c(ns)\rightarrow B_c(ms)+\pi+\pi\; (n>m)$, the
dominant mechanism is the two-gluon emission which eventually
hadronize into two pions. As is indicated in the
literature\cite{Guo:2006ai}, the other minor mechanisms such as the
subsequential pion emissions, may interfere with the amplitude
induced by the two-gluon emission mechanism to change the lineshape
of the differential width. But the total width is definitely
determined by the two-gluon emission, so that our numerical results
would give the decay width which can be measured in the future
experiments at LHCb and Z-factory. The effects induced by the
secondary mechanisms, may be measured at the Z-factory where a
cleaner environment can provide an opportunity to conduct accurate
measurements including the geometrical distribution of produced
pions.

In our calculations, we need to input several potential parameters
to calculate the masses of the excited states of $B_c(ns) \; (n>1)$.
Unlike for the charmonia $\psi$ and $\Upsilon$ families, lack of
data on their masses causes errors in our numerical results.

We employ the QCD-multipole expansion (QCDME) method to calculate
the corresponding widths. The specific potential forms for $B_c$
meson and hybrid $b\bar c g$ are selected  based on the lattice
results.
%We observe that several potential parameters are universal
%for $c\bar c$, $b\bar b$ systems, thus we suppose they are also
%applicable for the $b\bar c$ meson and by fitting the mass of the
%ground state $B_c$ we obtain the last parameter $\kappa$ for the
%$B_c$ system. With the potential we can calculate the spectra of the
%$B_c$ family\footnote{It is noted that like the case of the heavy
%quarkonia, only the estimate for spectra of low-lying states of
%$B_c$ is more accurate. }. Instead,  the parameters for hybrid
%$b\bar c g$ cannot be determined as we did for quarkonia because of
%lack of data, thus we set $\kappa'$ to be a free parameter which
%varies in a small region $0.4\sim 0.54$, namely between the
%parameters for $c\bar c$ and $b\bar b$.
With the potential and concerned parameters we find that the mass of
the ground hybrid $b\bar cg$ is $7.63\sim 7.64$ GeV as the quark
content $b\bar c$ is in a spin-singlet. The widths of
$B_c(3s)\rightarrow B_c(2s)+\pi\pi$ and $B_c(2s)\rightarrow
B_c(1s)+\pi\pi$ are not sensitive to the change of parameters within
the concerned range and it is similar to the cases for
$\Gamma(\Upsilon(ns)\rightarrow \Upsilon(ms)+\pi\pi)$. While
calculating the function $f^l_{if}$, we sum over the intermediate
states of appropriate quantum numbers and truncate the expansion at
$K=7$. Our calculation suggests that the decay widths of
$B_c(ns)\rightarrow B_c(ms)+\pi+\pi$ can reach a few of tens of KeV
which is of the same order as $\psi(ns)\rightarrow \psi(ms)+\pi\pi$
and $\Upsilon(ns)\rightarrow \Upsilon(ms)+\pi\pi$, therefore with
the luminosity of LHCb and the proposed Z-factory, there would be no
problem to make relatively accurate measurements on such pion
radiative decays.

$B_c$ mesons were marginally produced at the LEP-I, as the
luminosity of the proposed $Z^0$ can be at least three orders higher
than that of LEP-I, there should be sufficient data on $B_c$
available. Moreover, there is a large phase space for the excited
states $B_c(ns)\; (n>1)$, so their production rates are not
suppressed by the phase space and are similar to that for the ground
$B_c$, so that there should be a good chance to observe $B_c(ns)\to
B_c(ms)+\pi+\pi\; (n>m)$ at the $Z^0$ factory. As aforementioned,
the background at the $Z^0$ factory is relatively small and a clean
environment is expected.

It is believed that LHCb and even TEVATRON possess sufficient
database for observing such decays. In fact, $B_c$ production was
first observed at TEVATRON, and with the energy and luminosity of
LHCb, observation of $B_c(ns)\to B_c(ms)+\pi+\pi\; (n>m)$ is
definitely feasible. However, on other aspect, both LHC and TEVATRON
are hadron colliders, so the background is much messier. With the
efforts of the experts including theorists and experimentalists, it
is already possible to clearly distinguish the signal from the
background. Definitely, high quality generators are necessary for
analyzing all possible sources of background \cite{Chang-generator}.
By contrary, the background at the $Z^0$ factory is not so serious,
i.e. the QCD contamination is relatively alleviated, even though it
still exists. Detailed analysis on the possible background is a
rather difficult job and usually is done by experts. When preparing
the manuscript, we have consulted with our experimental colleagues
about the possibility of observing such decays and analysis on the
background, and here we can only make a very rough discussion.

Therefore, for getting a better understanding of $B_c$ meson and its
excited states, the $Z^0$ factory is definitely superior to the
hadron colliders.

No doubt, the present work is still a primary effort to find the
excited states of $B_c$ and study their structures, as well as that
for the hybrid states $|b\bar c g>$. One can be convinced that the
order of magnitude of the numerical results is trustworthy, so that
it is optimistic that measurements on such processes at LHCb and
even the proposed Z-factory can be conducted. When the data are
available, we will be able to further investigate the structure of
the $B_c$ family and identify the mechanism(s) which governs the
transitions. Then more precise theoretical works will be needed.\\

\section*{Acknowledgements}

We benefit greatly from discussions with C.H. Chang and Y.N. Gao.
This work is supported by the National Natural Science Foundation of
China (NNSFC) and the special grant for the PH.D program of the
Chinese Education Ministry. One of us (Ke) is also partly supported
by the special grant for new faculty from Tianjin University.
\\


\begin{thebibliography}{99}
%\cite{Chang:1991bp}
\bibitem{Chang}
  C.~Chang and Y.~Chen,
  %``The B(c) and anti-B(c) mesons accessible to experiments through Z0 bosons
  %decay,''
  Phys.\ Lett.\  B {\bf 284}, 127 (1992);
  %%CITATION = PHLTA,B284,127;%%
  %``The Production of B(c) or anti-B(c) meson associated with two heavy quark
  %jets in Z0 boson decay,''
  Phys.\ Rev.\  D {\bf 46}, 3845 (1992)
  [Erratum-ibid.\  D {\bf 50}, 6013 (1994)];
  %%CITATION = PHRVA,D46,3845;%%
  %``The hadronic production of the B(c) meson at Tevatron, CERN LHC and SSC,''
  Phys.\ Rev.\  D {\bf 48} (1993) 4086.
  %%CITATION = PHRVA,D48,4086;%%

%\cite{Cheung:1999ir}
\bibitem{Cheung:1999ir}
  K.~Cheung,
  %``$B_c$ meson production at the Tevatron revisited,''
  Phys.\ Lett.\  B {\bf 472}, 408 (2000)
  [arXiv:hep-ph/9908405];
  %%CITATION = PHLTA,B472,408;%%
%\cite{Wester:2005ex}
  W.~Wester  [CDF and D0 Collaborations],
  %``$B_c$ at the Tevatron,''
  Nucl.\ Phys.\ Proc.\ Suppl.\  {\bf 156}, 240 (2006).
  %%CITATION = NUPHZ,156,240;%%

\bibitem{Chang1} C. Chang et al. Phys.Rev.{\bf D64} (2001)
014003;Commun.Thor.Phys.{\bf 35} (2001) 51; V. Kiselev et al.
Nucl.Phys.{\bf B585} (2000) 353; A. Anisimov et al. Phys.Lett.{\bf
B452} (1999) 129; M. Beneke and G. Buchalla, Phys.Rev.{\bf D53}
(1996) 4991.






\bibitem{Gottfried} K. Gottfried, Phys. Rev. Lett. {\bf 40}, 598(1978).

 \bibitem{YK1} Y. Kuang and T.  Yan, Phys. Rev. {\bf D 24}, 2874(1981).

 \bibitem{K2} Y. Kuang , Front. Phys. China {\bf 1}, 19(2006).
  \bibitem{Y1} T. Yan, Phys. Rev. {\bf D 22}, 1652(1980).
 \bibitem{K1} Y.  Kuang, Y. Yi and B. Fu, Phys. Rev. {\bf D 42}, 2300(1990).
 \bibitem{soft} L. Brown and R. Cahn, Phys. Rev. Lett. {\bf 35}, 1(1975).
 \bibitem{Isgur} N. Isgur and J. Paton, Phys. Rev. {\bf D 31}, 2910(1985).
 \bibitem{BT}W. Buchm$\ddot{\rm u}$ller and H. Tye, Phys. Rev. Lett. {\bf 44}, 850(1980).

 \bibitem{Ke2007} H. Ke, J. Tang, X. Hao and X. Li, Phys. Rev. {\bf D76},
 074035(2007).


\bibitem{Swanson} E. Swanson and A. Szczepaniak, Phys. Rev. {\bf D 59}, 014035(1999).
\bibitem{Allen} T.  Allen, M. Olsson and S. Veseli,  Phys. Lett. {\bf B 434},
110(1998).
 \bibitem{cornell} E. Eichten, K. Gottfried, T. Kinashita, K. Lane and T. Yan
 , Phys. Rev. {\bf D 17}, 3090 (1978); $ibid$, {\bf D 21}, 203(1980).

 \bibitem{ss} T. Barnes, S. Godfrey and E. Swanson , Phys. Rev. {\bf D 72}, 054026(2005) .




\bibitem{PDG}
  C.~Amsler {\it et al.}  [Particle Data Group],
  %``Review of particle physics,''
  Phys.\ Lett.\  B {\bf 667}, 1 (2008);  V.~Abazov {\it et al.}  [D0 Collaboration],
  %``Observation of the $B_c$ Meson in the Exclusive Decay $B_c \to J/\psi
  %\pi$,''
  Phys.\ Rev.\ Lett.\  {\bf 101}, 012001 (2008)
  [arXiv:0802.4258 [hep-ex]];  %%CITATION = PRLTA,101,012001;%%
 M.~Spezziga  [CDF Collaboration],
  %``Studies of the $B_c$ meson at CDF,''
  Nucl.\ Phys.\ Proc.\ Suppl.\  {\bf 164}, 149 (2007)
  [arXiv:hep-ex/0511007].
  %%CITATION = NUPHZ,164,149;%%



\bibitem{Choi2009} H. Choi and C. Ji, arXiv:0903.0455[hep-ph].

\bibitem{Colangelo2000} P. Colangelo and F. DeFazio, Phys. Rev. {\bf D 61},
034012(2000); M. Baker, J. Ball and F. Zachariasen, Phys. Rev.
{\bf D 45}, 910(1992).

\bibitem{Guo:2006ai}
  F.~Guo, P.~Shen, H.~Chiang and R.~Ping,
  %``On the structure of the pi pi invariant mass spectra of the Upsilon(4S) -->
  %Upsilon(1S, 2S) pi+ pi-,''
  Phys.\ Lett.\  B {\bf 658}, 27 (2007)
  [arXiv:hep-ph/0601120];
  %%CITATION = PHLTA,B658,27;%%
%\cite{Guo:2006fv}
  F.~Guo, P.~Shen and H.~Jiang,
  %``The pi pi phase shifts from psi' --> J/psi pi+ pi- decays,''
  High Energy Phys.\ Nucl.\ Phys.\  {\bf 29}, 892 (2005)
  [arXiv:hep-ph/0601082].
  %%CITATION = KNWLD,29,892;%%

\bibitem{Chang-generator} C.~H.~Chang, J.~X.~Wang and X.~G.~Wu,
  %``BCVEGPY2.0: A upgrade version of the generator BCVEGPY with an addendum
  %about hadroproduction of the P-wave B/c states,''
  Comput.\ Phys.\ Commun.\  {\bf 174}, 241 (2006)
  [arXiv:hep-ph/0504017].
  %%CITATION = CPHCB,174,241;%%



\end{thebibliography}
\end{document}